\documentclass[journal]{IEEEtran}%
\usepackage{amsmath}
\usepackage{amsfonts}
\usepackage{amssymb}
\usepackage{graphicx}%
\begin{document}

\title{Nonlinear Damping in Nanomechanical Beam Oscillator}
\author{Stav Zaitsev,~\IEEEmembership{Student Member,~IEEE,} Ronen Almog, Oleg
Shtempluck and Eyal Buks\thanks{The authors are with the Electrical
Engineering Department, Technion, Israel Institute of Technology.}}
\maketitle

\begin{abstract}
We investigate the impact of nonlinear damping on the dynamics of a
nanomechanical doubly clamped beam. The beam is driven into nonlinear regime
and the response is measured by a displacement detector. For data analysis we
introduce a nonlinear damping term to Duffing equation. The experiment shows
conclusively that accounting for nonlinear damping effects is needed for
correct modeling of the nanomechanical resonators under study.

\end{abstract}


\IEEEpeerreviewmaketitle

\section{Introduction}

\PARstart{T}{he} field of micro-machining is forcing a profound redefinition
of the nature and attributes of electronic devices. This technology allows
fabrication of a variety of on-chip fully integrated sensors and actuators
with a rapidly growing range of applications. In many cases it is highly
desirable to shrink the size of mechanical elements down to the nano-scale
\cite{Roukes_NEMS_01, Roukes_NEMS_00}. This allows enhancing the speed of
operation by increasing the frequencies of mechanical resonances and improving
their sensitivity as sensors. Moreover, as devices become smaller, their power
consumption goes down and their cost can be significantly lowered. Some key
applications of nanoelectromechanical systems (NEMS) technology include
magnetic resonance force microscopy (MRFM) \cite{Sidles_et_al_95,
Rugar_et_al_04} and mass-sensing \cite{Ekinci_et_al_04}. Further
miniaturization is also motivated by the quest for mesoscopic quantum effects
in mechanical systems \cite{Blencowe_04, Knobel&Cleland_03, Lahaye_et_al_04}.

A key property of systems based on mechanical oscillators is the rate of
damping. For example, in many cases the sensitivity of NEMS sensors is limited
by thermal fluctuation which is related to damping via the fluctuation
dissipation theorem. In general, a variety of different physical mechanisms
can contribute to damping, including bulk and surface defects, thermoelastic
damping, nonlinear coupling to other modes, phonon-electron coupling, clamping
loss etc. Identifying experimentally the contributing mechanisms in a given
system can be highly challenging, as the dependence on a variety of parameters
has to be examined systematically. Nanomechanical systems suffer from low
quality factors $Q$ relative to their macroscopic\ counterparts
\cite{Roukes_NEMS_00}. This behavior suggests that damping in nanomechanical
devices is dominated by surface properties, since the relative number of atoms
on the surface or close to the surface increases as device dimensions
decrease. This point of view is also supported by some
experiments~\cite{Yasumura_et_al_00, Ono_et_al_03}. However, very little is
currently known about the underlying physical mechanisms contributing to
damping in these devices.

In the present paper we study damping in a nanomechanical oscillator operated
in the nonlinear regime.\ Nonlinear effects are of great importance for
nanomechanical devices. The relatively small applied forces needed for driving
a nanomechanical oscillator into the nonlinear regime is usually easily
accessible.\ Thus, a variety of useful applications such as frequency
synchronization, frequency mixing and conversion, parametric and
intermodulation amplification \cite{Almog_et_al_06a}, mechanical noise
squeezing \cite{Almog_et_al_06b}, and enhanced sensitivity mass detection
\cite{Buks&Yurke_06} can be implemented by applying modest driving
forces.\ Moreover, monitoring the displacement of a nanomechanical resonator
oscillating in the linear regime may be difficult when a displacement detector
with high sensitivity is not available. Thus, in many cases the nonlinear
regime is the only useful regime of operation. However, to optimize the
properties of NEMS devices operating in the nonlinear regime it is important
to characterize the effect of damping in this regime.

The effect of nonlinear damping for the case of strictly dissipative force,
being proportional to the velocity to the p'th power, on the response and
bifurcations of driven Duffing \cite{Ravindra&Mallik_94a, Ravindra&Mallik_94b,
Trueba_et_al_00, Baltanas_et_al_01} and other types of nonlinear oscillators
\cite{Nayfeh_Mook_book_95, Sanjuan_99, Trueba_et_al_00} have been studied
extensively. For the present case we consider a Duffing oscillator having
nonlinear damping force proportional to the velocity cubed. As will be shown
below, this approach is equivalent to the case where damping nonlinearity
proportional to the velocity multiplied by the displacement squared is
considered (see Ref. \cite{Lifshitz&Cross_03}). We have recently studied a
closely related problem of a nonlinear stripline superconducting
electromagnetic oscillator \cite{Yurke&Buks_05, Buks&Yurke_05}, where
nonlinear damping was taken into account. With some adjustments, these results
are implemented for the case of a nanomechanical nonlinear oscillator. To
determine experimentally the rate of nonlinear damping, as well as the Kerr
constant and other important\ parameters, we measure the response near the
resonance in the nonlinear regime \cite{Buks&Roukes_01a, Buks&Roukes_01b}.
Measuring these parameters under varying conditions provides important
insights into the underlying physical mechanisms.

\section{Experimental setup}

\begin{figure}
[h]
\begin{center}
\includegraphics[
height=1.9346in,
width=2.879in
]%
{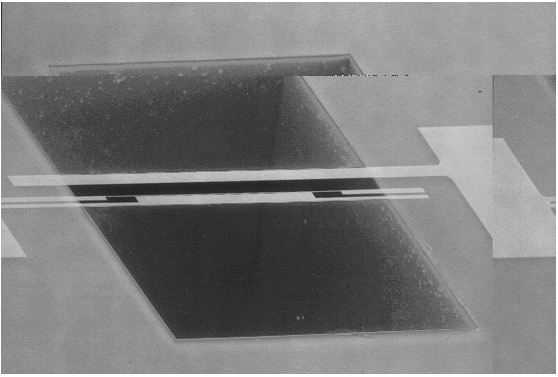}%
\caption{The device consists of a narrow cantilever beam (length
200$\operatorname{\mu m}$, width 1-0.25$\operatorname{\mu m}$ and thickness
0.2$\operatorname{\mu m}$) and wide electrode. The excitation force is applied
as voltage between the beam and the electrode.}%
\label{F_device}%
\end{center}
\end{figure}

For the experiments we employ nanomechanical oscillators in the form of doubly
clamped beams made of PdAu (see Fig. \ref{F_device}). The bulk nano-machining
process used for sample fabrication is similar to the one described in Ref.
\cite{Buks&Roukes_01b}. The dimensions of the beams are length 100-200$%
\operatorname{\mu m}%
$, width 0.25-1$%
\operatorname{\mu m}%
$ and thickness 0.2$%
\operatorname{\mu m}%
$, and the gap separating the beam and the electrode is 5$%
\operatorname{\mu m}%
$. Measurements of mechanical properties are done \textit{in-situ} a scanning
electron microscope, where the imaging system of the microscope is employed
for displacement detection \cite{Buks&Roukes_01b}. Some of the samples were
also measured using an optical displacement detection system described
elsewhere \cite{Almog_et_al_06b}. Driving force is applied to the beam by
applying a voltage to the nearby electrode. With a relatively modest driving
force the system is driven into the region of nonlinear oscillations
\cite{Buks&Roukes_01b, Buks&Roukes_02}.%

\section{Equation of motion}

We excite the system close to its linear fundamental mode. Ignoring all higher
modes allows us to describe the dynamics using a single degree of freedom $x$.

The nonlinear equation of motion is
\begin{equation}
m\ddot{x}+2b_{1}\dot{x}+k_{1}x+b_{3}\dot{x}^{3}+k_{3}x^{3}=-\frac
{\mathrm{d}\mathcal{E}_{cap}}{\mathrm{d}x}, \label{E_Newton_eq}%
\end{equation}
where $m$ is the effective mass of the fundamental mode, $\mathcal{E}%
_{cap}=C(x)V^{2}/2$ is the capacitance energy, $C(x)=C_{0}/(1-x/d)$ is the
displacement dependent capacitance$,\ d$ is the gap between the electrode and
the beam, $b_{1}$ is the linear damping constant, $b_{3}$ is the nonlinear
damping constant, $k_{1}$ is the linear spring constant and $k_{3}$ is the
nonlinear (Kerr) spring constant.

The applied voltage is composed of large DC and small AC components
$V(t)=V_{DC}+v\cos(\omega t)$ where $v$ is constant and $v\ll V_{DC}$. Thus
the equation of motion reads%
\begin{multline}
(1-x/d)^{2}(\ddot{x}+2\gamma_{1}\dot{x}+\omega_{0}^{2}x+\gamma_{3}\dot{x}%
^{3}+\alpha_{3}x^{3})\label{E_eq_of_motion}\\
=F\left(  1+2f\cos(\omega t)\right)  ,
\end{multline}
where $\omega_{0}=\sqrt{k_{1}/m}$, $\gamma_{1}=b_{1}/m=\omega_{0}/2Q$
($Q$\ being the mechanical quality factor), $\gamma_{3}=b_{3}/m$, $\alpha
_{3}=k_{3}/m$, $F=C_{0}V_{DC}^{2}/2md$ and $f=v/V_{DC}$.

\section{Multiple scales approximation}

We use the standard multiple scales method to solve Eq. (\ref{E_eq_of_motion})
\cite{Nayfeh_Mook_book_95, Nayfeh_book_81}. The harmonic excitation frequency
is assumed to be close to the primary resonance%
\[
\omega=\omega_{0}+\sigma,
\]
where $\sigma\ll\omega_{0}$\ is a small detuning parameter. We also assume the
linear damping coefficient $\gamma_{1}$, both coefficients of nonlinear terms,
$\gamma_{3}$, $\alpha_{3}$, and $1/d$ to be small. Keeping terms up to the
first order in the small parameters leads to the following form of the
solution for $x(t)$
\begin{equation}
x(t)=\frac{F}{\omega_{0}^{2}}+\left(  A\left(  t\right)  e^{i\omega_{0}%
t}+c.c.\right)  , \label{E_x0}%
\end{equation}
where the first term in the right-hand side is a constant displacement due to
electrostatic attractive force, and $A\left(  t\right)  $ is slowly varying
envelope (on the time scale of $1/\omega_{0}$). The differential equation for
$A\left(  t\right)  $ in this approximation is given by%
\begin{multline}
2\omega_{0}\left[  i\frac{\mathrm{d}A}{\mathrm{d}t}+\left(  i\gamma_{1}%
+\Delta\omega_{0}\right)  A\right] \label{E_A_evolution}\\
+3\left(  \alpha_{3}+i\gamma_{3}\omega_{0}^{3}\right)  A^{2}A^{\ast
}=Ffe^{i\sigma t},
\end{multline}
where $\Delta\omega_{0}=\left(  3\alpha_{3}F/\omega_{0}^{4}-2/d\right)
F/2\omega_{0}$ is a small correction to the linear resonance frequency
$\omega_{0}$.

The solution for $A\left(  t\right)  $\ can be represented as
\begin{equation}
A\left(  t\right)  =ae^{i\left(  \phi+\Delta\omega_{0}t\right)  }, \label{E_A}%
\end{equation}
where $a$\ and $\ \phi$\ are real. Substituting Eq. (\ref{E_A}) into Eq.
(\ref{E_A_evolution}) and separating real and imaginary parts one finds\
\begin{subequations}
\label{E_sec}%
\begin{align}
-2\omega_{0}a\frac{\mathrm{d}\phi}{\mathrm{d}t}+3\alpha_{3}a^{3}  &
=Ff\cos\left(  \phi-\Delta\omega t\right)  ,\label{E_sec_real}\\
-2\omega_{0}\left(  \frac{\mathrm{d}a}{\mathrm{d}t}+\gamma_{1}a\right)
-3\gamma_{3}\omega_{0}^{3}a^{3}  &  =Ff\sin\left(  \phi-\Delta\omega t\right)
, \label{E_sec_img}%
\end{align}
where $\Delta\omega=\sigma-\Delta\omega_{0}$\ is the excitation frequency
detuning from the shifted resonance frequency $\omega_{0}+\Delta\omega_{0}$.
In the steady state $a$\ and $\phi-\Delta\omega t$\ are constant and the
following equation for the steady state response amplitude $a$\ can be derived
from Eq. (\ref{E_sec})
\end{subequations}
\begin{multline}
9\left(  \alpha_{3}^{2}+\gamma_{3}^{2}\omega_{0}^{6}\right)  a^{6}%
+12\omega_{0}\left(  \gamma_{1}\gamma_{3}\omega_{0}^{3}-\Delta\omega\alpha
_{3}\right)  a^{4}\label{E_sec_a^2}\\
+4\omega_{0}^{2}\left(  \Delta\omega^{2}+\gamma_{1}^{2}\right)  a^{2}%
-F^{2}f^{2}=0,
\end{multline}

Equation of the same form was obtained in Ref. \cite{Yurke&Buks_05}, where a
superconducting oscillator having Kerr nonlinearity in addition to nonlinear
damping was considered. All subsequent analysis is thus based on Ref.
\cite{Yurke&Buks_05}.

\begin{figure}
[h]
\begin{center}
\includegraphics[
height=3.3996in,
width=2.367in
]%
{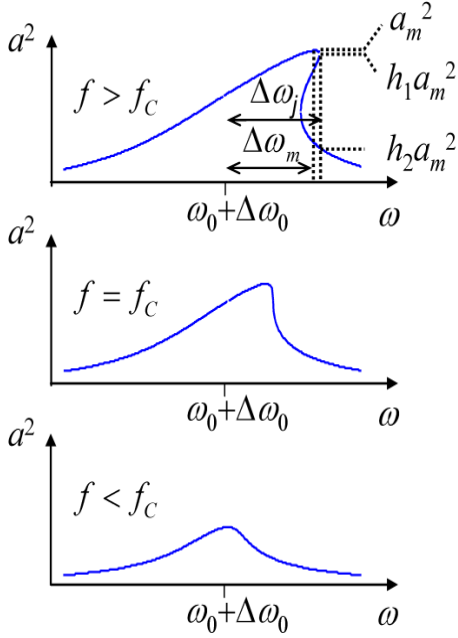}%
\caption{Steady state solutions under different excitation amplitudes $f$. In
case $f<f_{c}$ only one real solution exists, no bistability is possible. In
case $\ f=f_{c}$ the system is on the edge of bistability, and one point
exists where $a^{2}\ $vs.\ $\omega$ has infinite slope. In case $\ f>f_{c}$
the system is in bistable regime having three real solutions over some range
of frequencies. Two of these solutions are stable. }%
\label{F_Duffing}%
\end{center}
\end{figure}

When $\gamma_{3}$ is sufficiently small the solutions of Eq. (\ref{E_sec_a^2})
behave very much like the ordinary Duffing equation solutions to which Eq.
(\ref{E_Newton_eq}) reduces to when $b_{3}=0$ (see Fig. \ref{F_Duffing}).%

Interestingly enough, equations similar to Eq. (\ref{E_A_evolution}) and Eq.
(\ref{E_sec_a^2}) arise when the damping nonlinearity is considered to be
proportional to velocity multiplied by the displacement squared (instead of
velocity cubed)%
\[
m\ddot{x}+2b_{1}\dot{x}+k_{1}x+b_{3}x^{2}\dot{x}+k_{3}x^{3}=-\frac
{\mathrm{d}\mathcal{E}_{cap}}{\mathrm{d}x}.
\]
Substituting $\gamma_{1}$\ by $\left(  \gamma_{1}+\gamma_{3}F^{2}/2\omega
_{0}^{4}\right)  $\ and $\gamma_{3}$\ by $\gamma_{3}\left(  1+2F/\omega
_{0}^{2}\right)  /3\omega_{0}^{2}$\ in Eq. (\ref{E_A_evolution}) and Eq.
(\ref{E_sec_a^2}) gives the correct relations for this case. Therefore, the
behavior for these two cases is similar near the resonance frequency.

\section{Special points}

Referring to Fig. \ref{F_Duffing} we define some points in the $a^{2}$ vs.
$\omega$ curves which we use in experimental data analysis.

The first point is the maximum response, shifted by $\Delta\omega_{m}$\ from
$\omega_{0}+\Delta\omega_{0}$ and having the amplitude $a_{m}$.
Differentiating Eq. (\ref{E_sec_a^2}) with respect to $\Delta\omega$ and
demanding $\mathrm{d}\left(  a^{2}\right)  /\mathrm{d}\Delta\omega=0$ yields%

\begin{equation}
a_{m}^{2}=\frac{2\omega_{0}\Delta\omega_{m}}{3\alpha_{3}}. \label{E_am^2}%
\end{equation}

Another point of special interest is the point where the jump in amplitude
occurs and therefore the condition $\mathrm{d}\Delta\omega/\mathrm{d}\left(
a^{2}\right)  =0$ must be satisfied. Applying this condition to Eq.
(\ref{E_sec_a^2}) yields%

\begin{multline}
27\left(  \alpha_{3}^{2}+\gamma_{3}^{2}\omega_{0}^{6}\right)  a^{4}%
+24\omega_{0}\left(  \gamma_{1}\gamma_{3}\omega_{0}^{3}-\Delta\omega\alpha
_{3}\right)  a^{2}\label{E_jump_points}\\
+4\omega_{0}^{2}\left(  \Delta\omega^{2}+\gamma_{1}^{2}\right)  =0.
\end{multline}

\begin{figure}
[ptb]
\begin{center}
\includegraphics[
height=3.0485in,
width=2.744in
]%
{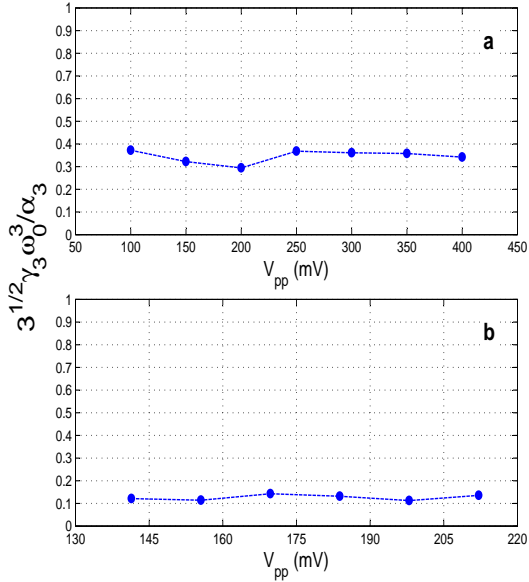}%
\caption{Experimental results for $p=\sqrt{3}\gamma_{3}\omega_{0}^{3}%
/\alpha_{3}$ vs. peak-to-peak excitation amplitude $V_{pp}$ . (a)
200$\operatorname{\mu m}$ long beam with fundamental mode occuring at
$f_{0}=123.2$ $\operatorname{kHz}$ and $Q=7200$. (b) 125$\operatorname{\mu m}$
long beam with fundamental mode occuring at $f_{0}=524.6$ $\operatorname{kHz}$
and $Q=2100$.}%
\label{F_p_plot}%
\end{center}
\end{figure}

Eq. (\ref{E_jump_points}) has a single real solution at the point of critical
frequency $\Delta\omega_{c}$ and critical amplitude $a_{c}$, where the system
is on the edge of bistability. This point is defined by two conditions
\begin{subequations}
\begin{gather*}
\frac{\mathrm{d}\Delta\omega}{\mathrm{d}\left(  a^{2}\right)  }=0,\\
\frac{\mathrm{d}^{2}\Delta\omega}{\mathrm{d}(a^{2})^{2}}=0.
\end{gather*}

In general, $\gamma_{3}$ is positive but $\alpha_{3}$ can be either positive
(hard spring) or negative (soft spring). In our experiment $\alpha_{3}>0$. By
applying these conditions one finds%

\end{subequations}
\begin{subequations}
\label{E_critical}%
\begin{gather}
\Delta\omega_{c}=\frac{\gamma_{1}}{\sqrt{3}}\frac{p+3}{1-p}%
,\label{E_critical_freq}\\
a_{c}^{2}=\frac{4}{3\sqrt{3}}\frac{\gamma_{1}\omega_{0}}{\alpha_{3}}\frac
{1}{1-p}, \label{E_critical_a}%
\end{gather}
where $p=\sqrt{3}\gamma_{3}\omega_{0}^{3}/\alpha_{3}$. The driving force at
this critical point is denoted in Fig. \ref{F_Duffing} as $f_{C}$. Note that
bistable region is accessible only when $p<1.$

\section{Experimental data and results}

A typical measured response of the fundamental mode of a 200$%
\operatorname{\mu m}%
$ (125$%
\operatorname{\mu m}%
$)\ long beam occurring at $f_{0}=123.2%
\operatorname{kHz}%
$ ($f_{0}=524.6%
\operatorname{kHz}%
$) measured with $V_{DC}=20%
\operatorname{V}%
$ and varying excitation amplitude is seen in Fig. \ref{F_raw_data}(a) (Fig.
\ref{F_raw_data}(b)). We derive the value of $\gamma_{1}=\omega_{0}/2Q$ from
the linear response at low excitation amplitude and find $Q=7200$ ($Q=2100$).%

\begin{figure}
[h]
\begin{center}
\includegraphics[
height=3.6158in,
width=3.3978in
]%
{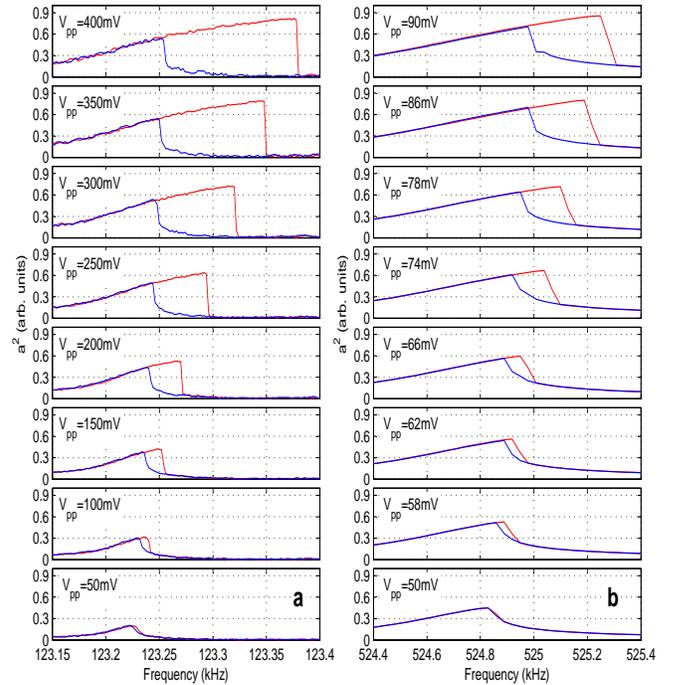}%
\caption{Measured response vs. frequency shown for both upward and downward
frequency sweeps with $V_{DC}=20$V and with varying peak-to-peak excitation
amplitude $V_{pp}$. (a) 200$\operatorname{\mu m}$ long beam with fundamental
mode occuring at $f_{0}=123.2$ $\operatorname{kHz}$. (b)
125$\operatorname{\mu m}$ long beam with fundamental mode occuring at
$f_{0}=524.6$ $\operatorname{kHz}$.}%
\label{F_raw_data}%
\end{center}
\end{figure}

Our displacement detector is highly nonlinear, introducing thus a significant
distortion in the measured response. In order to minimize the resultant
inaccuracies, we employ the following method to extract the nonlinear parameters.

In general, the sum of the three solutions for $a^{2}$ at any given frequency
can be found from Eq. (\ref{E_sec_a^2}). This is employed for the jump point
at $\omega_{0}+\Delta\omega_{j}$ seen in Fig. \ref{F_Duffing}. Using Eq.
(\ref{E_am^2}) to calibrate the measured response at this jump point one has
\end{subequations}
\begin{subequations}
\[
(2h_{1}+h_{2})\frac{2\omega_{0}\Delta\omega_{m}}{3\alpha_{3}}=-\frac
{4\omega_{0}\left(  \gamma_{1}\gamma_{3}\omega_{0}^{3}-\Delta\omega_{j}%
\alpha_{3}\right)  }{3\left(  \alpha_{3}^{2}+\gamma_{3}^{2}\omega_{0}%
^{6}\right)  },
\]
or
\end{subequations}
\begin{equation}
\left(  2h_{1}+h_{2}\right)  \Delta\omega_{m}\left(  \frac{p^{2}}{3}+1\right)
+2\left(  \gamma_{1}\frac{p}{\sqrt{3}}-\Delta\omega_{j}\right)  =0,
\label{E_p_quadratic_eq}%
\end{equation}
where $h_{1}$ and $h_{2}$ are defined in Fig. \ref{F_Duffing}. Due to the
frequency proximity between the maximum point and the jump point at
$\omega=\omega_{0}+\Delta\omega_{j}$ the inaccuracy of such a calibration is
small. Moreover, as long as excitation amplitude is high enough, $h_{2}$ is
much smaller than $h_{1}$ and even considerable inaccuracy in $h_{2}$
estimation will not have any significant impact. This equation can be used to
estimate $p$ for different excitation amplitudes. The results of applying Eq.
(\ref{E_p_quadratic_eq}) to experimental data\ from the two different beams
can be seen in Fig. \ref{F_p_plot}.%

Using this procedure we find $p\approx0.35$\ for the 200$%
\operatorname{\mu m}%
$\ long beam and $p\approx0.14$\ for the 125$%
\operatorname{\mu m}%
$\ long beam. These results are identical to the values of $p$ estimated using
Eq. (\ref{E_critical_freq}). Referring to Eq. (\ref{E_critical}) and Eq.
(\ref{E_jump_points}) we see that in our system the damping nonlinearity is
not negligible and has a measurable impact on both the amplitude and frequency
offset of the critical point, as well as on jump points in the bistable region.

To determine the value of $\alpha_{3}$ we measure the static deflection of the
beam's center as a function of an applied DC voltage $V_{DC}$. From a fit to
theory we find $\alpha_{3}=\omega_{0}^{2}\cdot0.092%
\operatorname{\mu m}%
{}^{-2}$\ for the 200$%
\operatorname{\mu m}%
$\ long beam.

\section{Conclusions}

In this work we have demonstrated conclusively that nonlinear damping in
nanomechanical doubly-clamped beam oscillator may play an important role. The
method presented in this paper may allow a systematic study of nonlinear
damping in nano-mechanical oscillators, which may help revealing the
underlying physical mechanisms.

\section*{Acknowledgment}

We would like to thank B. Yurke, O. Gottlieb and R. Lifshitz for many fruitful
discussions. This work was partially supported by Intel Corporation, US-Israel
binational foundation and by the Israeli Ministry of Science.

\bibliographystyle{IEEEtran}
\bibliography{GENERAL}

\end{document}